\documentclass[useAMS,usenatbib]{mn2e}
\usepackage{graphicx} 
\usepackage{tabularx}

\title[A dynamical stability study of Kepler Circumbinary Planetary systems with one planet]
  {A dynamical stability study of Kepler Circumbinary Planetary systems with one planet}
\author[C. E. Chavez et al.]
  {C. E. Chavez,$^1$\thanks{CARLOS.CHAVEZPCH@uanl.edu.mx}
  N. Georgakarakos,$^{2, 6}$ S. Prodan,$^3$ M. Reyes-Ruiz,$^5$
  H. Aceves$^5$ 
  \newauthor 
  F. Betancourt,$^4$ E. Perez-Tijerina,$^4$ \\
  $^1$Universidad Auton\'oma de Nuevo Le\'on, Facultad de Ingenieria Mec\'anica y El\'ectrica, San Nicol\'as de los Garza, N.L., M\'exico\\
  $^2$Higher Technological Educational Institute of Central Macedonia, Terma Magnesias, Serres 62124, Greece\\
  $^3$Dept. of Physics/CITA University of Toronto 60 St. George Street Toronto, ON, M5S 3H8, Canada\\
  $^4$Universidad Aut\'onoma de Nuevo Le\'on, Facultad de Ciencias F\'\i{}sico-Matem\'aticas, San Nicol\'as de los Garza, N.L., M\'exico.\\
  $^5$Universidad Nacional Aut\'onoma de M\'exico, Instituto de Astronom\'\i{}a de la UNAM, Ensenada 22860, B.C., M\'exico\\
  $^6$New York University Abu Dhabi, Saadiyat Island, Abu Dhabi, UAE}	
\date{Released 2014 Xxxxx XX}

\pagerange{\pageref{firstpage}--\pageref{lastpage}} \pubyear{2014}

\def\LaTeX{L\kern-.36em\raise.3ex\hbox{a}\kern-.15em
    T\kern-.1667em\lower.7ex\hbox{E}\kern-.125emX}

\begin{document}

\label{firstpage}

\maketitle

\begin{abstract}
To date, 17 circumbinary planets have been discovered.  In this paper, we focus our attention on the stability of the Kepler circumbinary planetary systems with only one planet, i.e. Kepler-16, Kepler-34, Kepler-35, Kepler-38, Kepler-64 and Kepler-413. In addition to their intrinsic interest, the study of such systems is an opportunity to test our understanding of planetary system formation and evolution around binaries.  The investigation is done by means of numerical simulations.  We perform numerical integrations of the full equations of motion of each system with the aim of checking the stability of the planetary orbit.  The investigation of the stability of the above systems consists of three numerical experiments.  In the first one we perform a long term (1Gyr) numerical integration of the nominal solution of the six Kepler systems under investigation. 
In the second experiment, we look for the critical semimajor axis of the six planetary orbits, and finally, in the third experiment, we construct two dimensional stability maps on the eccentricity-pericentre distance plane.
Additionally, using numerical integrations of the nominal solutions we checked if this solutions were close to the exact resonance.

\end{abstract}


\section{Introduction}
Before 1992, the only planets we knew were the ones of our own Solar System. To date, more than
one thousand and five hundred planets are known to revolve around stars other than the Sun. Many of the new
planetary systems exhibit widely characteristics from our Solar System (e.g. medium or high eccentricities
and the presence of Jupiter sized bodies close to the star) and consequently this has led to a major revision the planetary formation and evolution theories.  Modern observations give values of up to 70${\%}$ for the frequency of multiple stellar systems in our Galaxy (Kiseleva-Eggleton and  Eggleton 2001 and references therein), which makes the study of the process of planet formation and evolution in such systems, an important field of study.  

Nowadays, over seventy binary systems have been discovered harbouring planets (e.g. see Roell et al. 2012) and even more are expected to be found.  The interest of the scientific community, regarding, besides planet formation, various other issues related to planets in stellar binaries, such as habitability (e.g. Eggl et al. 2012, Eggl et al. 2013, Kane 
\& Hinkel 2013), has greatly increased.  Although many questions still remain, significant effort has been channeled into formation studies in stellar binaries, which is also a test field for theories of planetary formation around single stars. 
Planet formation in binaries is currently a hotly debated topic that puzzles the scientific community,
especially as in some of the discovered systems the circumstellar planet is very close to its parent star (e.g. the ${\gamma}$ Cephei system, Hatzes et. al 2003, for which the planet semimajor axis is ${\sim}$ 2 AU and the companion star has semimajor axis ${\sim}$ 20 AU and eccentricity ${\sim}$ 0.36).  Paardekooper et al. (2012), Meschiari (2012), Pelupessy \& Portegies Zwart (2013), Thebault \& Haghighipour (2014) may provide the reader with relevant information regarding current issues of planet formation in stellar binaries. 

Most of the planets that have been discovered in stellar binaries are in circumstellar orbits. The first planet orbiting both stars of a stellar binary was found in the pulsar system PSR B1620-26 (Backer et al. 1993) with a  minimum mass between 1.5 and 3.5 M$_{J}$.  More recently, a system with two circumbinary planets was found to orbit HW Virginis (Lee, J. W. et al. 2009) and another pair of planets was detected around the eclipsing binary NN Serpentis (Qian et al. 2009, Beuermann et al. 2010).  A planet was found in another eclipsing binary, DP Leonis (Qian et al. 2010).  In the same year, a planet orbiting DT Virginis, also known as Ross 458 AB, was detected by direct imaging (Goldman et al. 2010, Burgasser et al. 2010).  At around the same time (Chavez et al. 2012), the light curve of the cataclysmic variable FS Aurigae was found to show the presence of a third body. In that study, it was found that the most plausible explanation to the observed variations in the light curve of the system was the presence of a third body on a wide orbit with a mass of 48 M$_{J}$, but a mass as small as 2 M$_{J}$ was not ruled out.  

The first circumbinary planet in the Kepler catalog was Kepler-16b (Doyle et al. 2011); this circumbinary planet was the first one to be confirmed by two different detection methods, i.e.  by the transit method and by RV measurememnts. As a consequence, its parameters are better constrained compared to the rest of the circumbinary planets and the same can be said for the rest of the circumbinary planets that have now been confirmed by Kepler.  A new planet was found to orbit NY Virginis in 2012 (Qian et al. 2012) and shortly after, another planet (possibly accompanied by an additional one) was reported to orbit the RR Cae system, an eclipsing binary containing white dwarfs (Qian et al. 2012b).  More circumbinary planets were found by Kepler: Kepler-34b and Kepler-35b (Welsh et al. 2012), Kepler-38b (Orosz et al. 2012a), Kepler-47b and c (Orosz et al. 2012b, Kostov et al. 2013), Kepler-64b (Kostov et al. 2013, Schwamb et al. 2013) and finally, the most recent of such Kepler discoveries has been Kepler-413b (Kostov et al. 2014). 

Previously, there has been some work regarding the stability of those systems.  Doyle et al. (2011) integrated numerically the best-fitting solution for Kepler-16b for two million years and they noticed no significant excursions in orbital distance that might lead to instability.  The stability of the Kepler-16 system was also investigated by Jaime et al. (2012).  In the context of determining areas of stable orbits around stellar binaries based on the concept of 'invariant loops' (e.g. Pichardo et al. 2005), they found that the Kepler-16b planet was lying in the stable region around the stellar binary, but not far away from the unstable area.  In addition, Popova and Shevchenko (2013) investigated the stability of the Kepler-16, systems by constructing stability maps based on the Lyapunov exponent concept and on an 'escape-collision' criterion, acording to which an orbit was classified as stable if the distance between the planet and one of the stars did not become less than 10$^{-3}$ AU or greater than 10$^{3}$AU.  They found that the planet Kepler-16b turns out to be just outside the chaotic area, among  some 'teeth' of chaotic motion corresponding to certain mean motion resonances. Additionally, they found similar results when applied their method to Kepler-34 and Kepler-35.

Welsh et al. (2012), along with the announcement of the discovery of Kepler-34b and Kepler-35b, performed a series of numerical integrations in order to assess the  stability of the two planetary systems.  Using the nominal orbital solutions, they performed
direct N-body integrations for 10 Myrs and found no indication of instability.   They also carried out 1000 simulations of systems
with different masses and orbital parameters for 1 Myrs, without finding, again, any sign of instability.  Moreover, they integrated an ensemble of a few thousand three-body systems, each consistent with the observed masses and orbital parameters, except
that they varied the semi-major axis of the planet with the aim of finding the minimum semi-major axis for which the system was not
flagged as unstable (i.e. change of the initial value of the planetary semi-major axis by more than 50\%) during the 10000 yr simulations.  Besides Kepler-34b and Kepler-35b, that was also done in the same article for Kepler-16b.
Regarding Kepler-38b, Orosz et al. (2012a) tested the stability of the system by performing direct N-body simulations and found the critical period to be 75 days. Kostov et al. (2013) and (2014) used the MEGNO method to investigate the stability of the Kepler-64 and the Kepler-413 systems respectively.  Carrying out numerical simulations with an integration time of 2700 yrs (which corresponded to ${5 \times 10^{4}}$ binary periods) for the former and 200000 days (which corresponded to about ${2 \times 10^{4}}$ binary periods) for the latter, and after they constructed dynamical maps, they concluded that the best-fit orbital parameters located the planets in a quasi-periodic region in (a, e) space, rendering the orbital solution to be plausible from a dynamical point of view.  Regarding Kepler-64, that conclusion seems to be confirmed by Schwamb et al. 2013, who carried out numerical integrations of the system using the best fit physical and orbital parameters and found the system to be stable on gigayear timescales, without giving any further information though.

In this paper, we focus our attention on the long term stability and dynamical evolution of the Kepler planetary systems with only one circumbinary planet.
The investigation is done by means of numerical simulations.  The paper is structured as follows: in section 2, we describe the method we use and the setup of our numerical experiments. The results of the various simulations conducted are presented in 
section 3. In section 4 we discuss the implications of our results and present concluding remarks.

\section{Numerical modelling and Method}
\label{sec:problem}
In this paper, we focus on the long term evolution and stability of the six single-planet circumbinary Kepler planetary systems.  In the context of the three body problem, we perform numerical integrations of the full equations of motion.  Our investigation is split into three numerical experiments.

In the first experiment, we used the nominal orbital solution of the six systems, as given in the relevant literature, and
we integrated their motion for 1 Gyr. This a reasonable amount of time to choose for our purposes as it is a considerable fraction of the time that the stars will remain on the main sequence and it is also much longer compared to almost all previous studies mentioned above, as most of the latter investigated the stability of the systems for a rather limited time interval ($\leq 10$ Myrs).  All
orbits were assumed coplanar, as the systems under investigation have a mutual inclination of the order of just a few degrees
and using a three dimensional model would have an insignificant effect on the outcome (e.g. see Doolin and Blundell 2011, Georgakarakos 2013).

The initial conditions for our simulations were derived from the orbital elements of the systems that were published in the relevant papers.  More specifically, for Kepler-16, Kepler-34 and Kepler-35, we used the data that appear in 'Supplementary Table 1' in 'Supplementary Information' of Welsh et al. (2012).  For Kepler-38, we obtained the values of all parameters from Tables 5, 6 and 7 that appear in Orosz et al. (2012a), using the column 'Best Fit' in all cases (for the planetary mass we used the maximum proposed value).  For the case of Kepler-64 we used two set of values since there are two published articles for this system. The first set appears in Tables 2 and 3 in Kostov et al. (2013) (again, for the planetary mass, we used the maximum proposed value), while the second set
appears in Tables 7 and 8 in Schwamb et al. (2013). Finally, for Kepler-413 we used the 'Best Fit' column in Tables 3 and 4 in Kostov et al. (2014). The mean longitudes $\lambda$ for Kepler-16, Kepler-34 and Kepler-35, are given in the corresponding articles, while for the case of Kepler-38, Kepler-64 (Kostov solution) and Kepler-413 the time of pericentre passage of the binary T$_{peri}$ is given and it can be used to calculate the mean longitude $\lambda$ by using Eqs. (2.39) and (2.53) from Murray and Dermott (1999).  For the case of the Schwamb solution of Kepler-64, we used the same time of pericentre passage of the binary that appears in Kostov et al. (2013), since that parameter does not appear in the corresponding article. In order to compute the missing mean longitudes, we used the fact that the times on which the planet transits the primary are given for all systems, therefore we took the time of the first observed primary transit and we evolved the inner binary from the its passage of the pericentre epoch to that given time. Since the planet has to transit the primary, it means that the Earth, the planet and the primary should be in the same line. The transit times for Kepler-38, Kepler-64 (Kostov), 
Kepler-64 (Schwamb), Kepler-413 were taken from Table 1 of Orosz et al. (2012), Table 3 of Kostov et al. (2013), Table 3 of Schwamb et al. (2013) and Table 5 from Kostov et al. (2014) respectively. 
Tables 1 and 2 give the orbital elements and masses of the systems under investigation.

\begin{table*}
\centering
\begin{minipage}{140mm}
  \caption{Masses and orbital elements for the stellar binary of each system. In the case of Kepler-64 we used two different sources for these elements, the first one being Kostov et al. 2013 (K) and the second one being Schwamb et al. 2013 (S); note that the node of the binary orbit does not appear in any case, since in all reference articles the relative nodal longitude is given.} 
 \label{tab:1}
 \begin{tabular}{@{}lrrrrrrrr@{}}
    \hline
    \hline
    System & ${M_{1}-M_{2} (M_{\odot})}$ & $a_b$ (AU) & $e_b$ & $i_b$ (deg) & $\omega_b$ (deg) & $\lambda_b$  (deg) & T$_{peri}$ (JD) \\
    \hline
    \hline
    Kepler-16\footnote{We use the 'Supplementary Table 1' in Supplementary Information' of Welsh et al. (2012)} & 0.6897-0.20255 & 0.22431 & 0.15944   & 90.3401 & -96.536  & 92.3520 & --\\
    Kepler-34 \textsuperscript{$a$} & 1.0479-1.0208 & 0.22882 & 0.52087 & 89.8584 & 71.4359  & 300.1970 & --\\
    Kepler-35 \textsuperscript{$a$} & 0.8876-0.8094 & 0.17617 & 0.1421 & 90.4238 & 86.5127 & 89.1784 & --\\
    Kepler-38\footnote{ We use Tables 5, 6 and 7 of Orosz et al. (2012a) using 'Best Fit'} & 0.949-0.249 & 0.1469 & 0.1032   & 89.265 & 268.680 & 62.96 & 2454971.66790\\
    Kepler-64 (K)\footnote{ We use Tables 2 and 3 of Kostov et al. (2013)} & 1.47-0.37 & 0.1769 & 0.204   & 87.59 & 214.3  & 159.6 & 2454973.862\\
    Kepler-64 (S)\footnote{ We use Tables 7 and 8 of Scwamb et al. (2013)} & 1.384-0.386 & 0.1744 & 0.2117 & 87.360 & 217.6  & 162.6 & 2454973.862\\
    Kepler-413\footnote{ We use the 'Best Fit' collumn in Tables 3 and 4 of Kostov et al. (2014)} & 0.820-0.5423 & 0.10148 & 0.0365 & 87.332 & 279.74  & 356.5 & 2454973.230 \\
       \hline
  \end{tabular}
  \end{minipage}
\end{table*}

\begin{table*}
\centering
\begin{minipage}{140mm}
 \caption{Planetary masses and orbital elements of each system. The node of the planet corresponds to the relative nodal longitude. For Kepler-64 (K) we assumed $\Omega_{p}=0$ since is not given in the article.} 
 \label{tab:2}
 \begin{tabular}{@{}lrrrrrrrr@{}}
    \hline
    \hline
    System & ${m_{p}}$ (${M_{J}}$) & $a_p$ (AU) & $e_p$ & $i_p$ (deg) & $\omega_p$ (deg) & $\Omega_p$ (deg) & $\lambda_p$  (deg)\\
    \hline
    \hline
    Kepler-16\footnote{ We use the 'Supplementary Table 1' in Supplementary Information' of Welsh et al. (2012)} & 0.333 &0.7048 & 0.00685  & 90.0322 & -41.2971 & 0.003 & 106.51 \\
    Kepler-34 \textsuperscript{$a$} & 0.22 &1.0896 & 0.182   & 90.355 & 82.0928 & -1.74 & 106.5\\
    Kepler-35 \textsuperscript{$a$} & 0.127 &0.60345 & 0.042   & 90.76 & 63.4349 & -1.24 & 136.4\\
    Kepler-38\footnote{ We use Tables 5 and 6 of Orosz et al. (2012a) using 'Best Fit'} &$<$0.384 &0.4644 & $<$0.032   & 89.446  & 32.8285 & -0.012 & 274.7\\
    Kepler-64 (K)\footnote{ We use Tables 2 and 3 of Kostov et al. (2013)} & $<$5  &0.642 & 0.1 & 90.0 & 105.0 & 0.0 & 264.43\\
    Kepler-64 (S)\footnote{ We use Tables 7 and 8 of Scwamb et al. (2013)} & $\leq$ 0.532  &0.634 & 0.0539 & 90.022 & 348.0 & 0.89 & 275.4\\
    Kepler-413\footnote{ We use the 'Best Fit' collumn in Tables 3 and 4 of Kostov et al. (2014)} & 0.21  &0.3553 & 0.1181 & 90.022 & 94.60 & 3.139& 283.1\\
       \hline
  \end{tabular}
    \end{minipage}
\end{table*}

In the second numerical experiment we searched for the critical planetary semi-major axis, i.e. the smallest planetary semi-major axis for which the planet would be on a stable orbit in the vicinity of the stellar binary,  just before the planetary orbit becomes unstable for the first time as we approach the binary.  Starting with the orbital
elements of each system the same as in Tables 1 and 2, we varied only the planetary semi-major axis with the aim of finding the value for which the planetary orbit would become unstable for the first time as we moved closer to the stellar binary.  These simulations were carried out for ${10^{5}}$ years (longer than any other similar study).  Making a reasonable choice for the initial semi-major axis, it was then reduced by a step of 0.01AU until the planetary orbit became unstable.  For each value of the semi-major axis, the planet was started at eight different initial positions, i.e. the initial planetary true anomaly was given values from ${0^{\circ}}$ to ${360^{\circ}}$ with a spacing of ${45^{\circ}}$.  If the planetary orbit became unstable at one or more of the initial positions, then, for that value of the semi-major axis, the planetary orbit was classified as unstable. 
%
%
%
%
%
\begin{figure}
\includegraphics[width=90mm,height=60mm,angle=0]{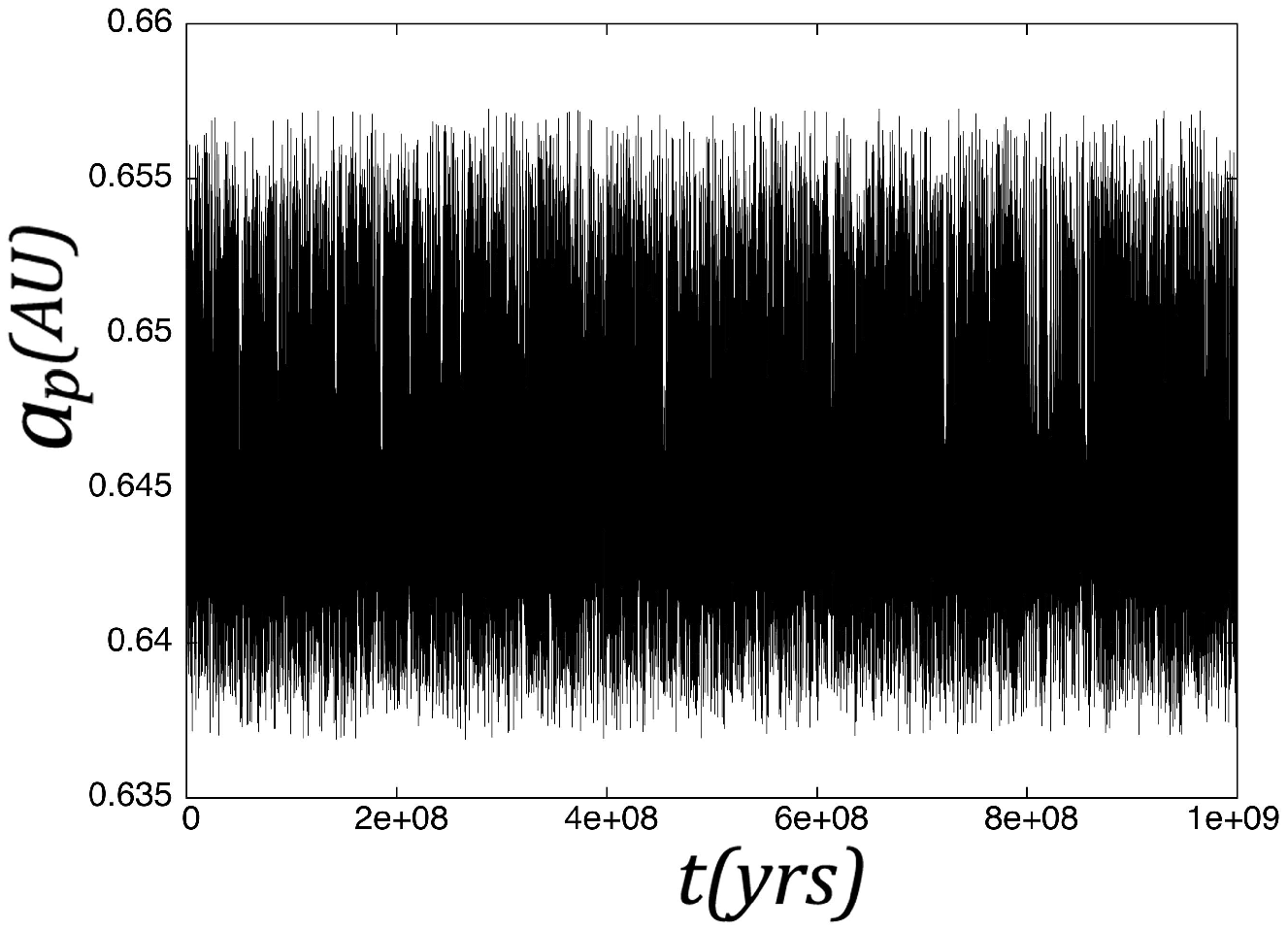}
\includegraphics[width=90mm,height=60mm,angle=0]{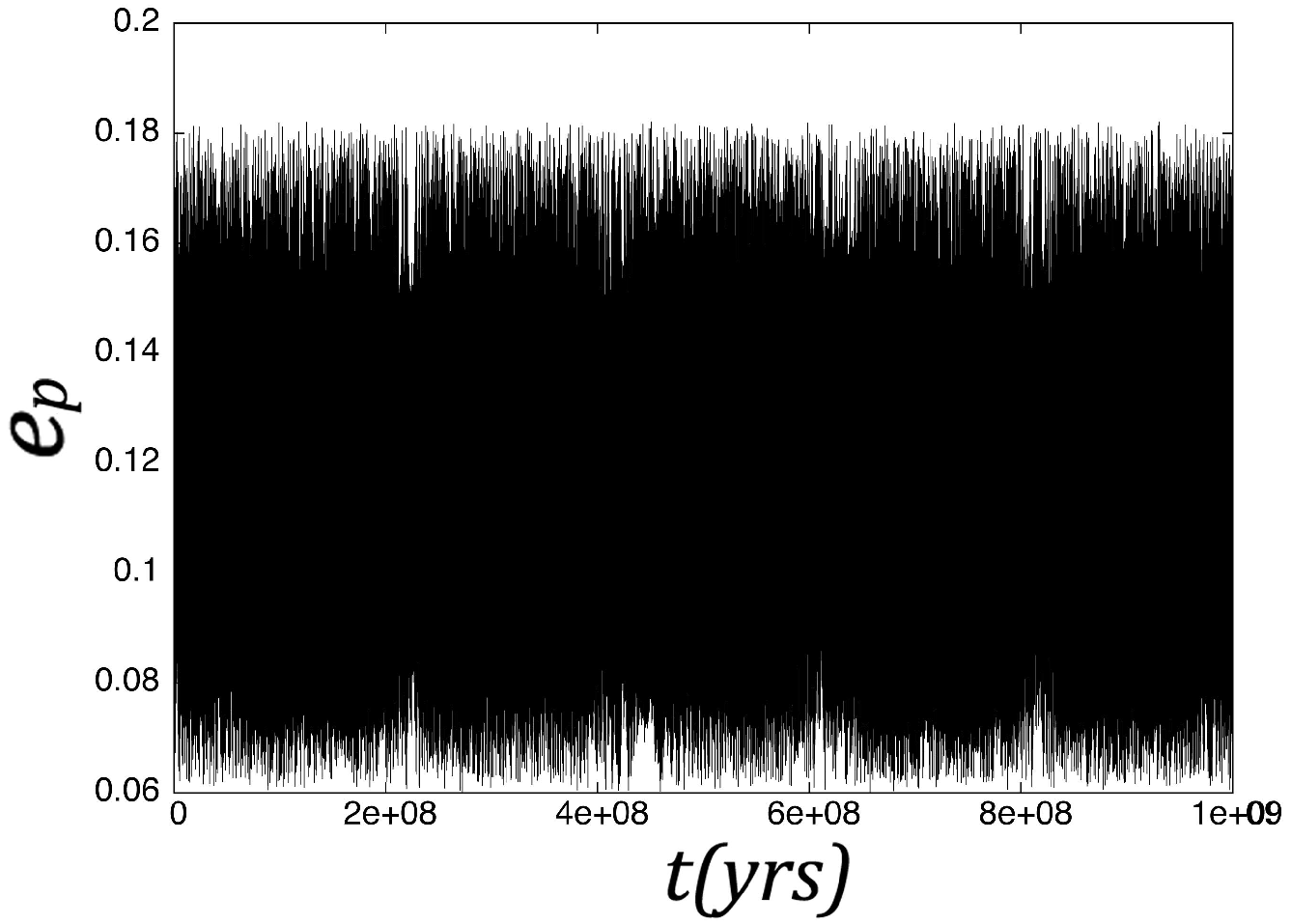}\\
\caption {Semi-major axis and eccentricity evolution for Kepler-64b based on Kostov et al. 2013.}
\label{fig5}
\end{figure}
%
%
%
%

Finally, in the third experiment, we searched for unstable areas on the ${(q_p,e_p)}$ plane (${e_p}$ being the initial planetary eccentricity and ${q_p}$ being the initial planetary pericentre distance) for each planetary system.  We constructed a grid of 101$\times$91 models, corresponding to the same number of ${(q_p,e_p)}$ points, with ${e_p}$ ranging from 0 to 0.9 and making a reasonable choice for our ${q_p}$ values by having a look at the stellar apocentre and planetary pericentre distances. As in the search for the critical semi-major axis of the nominal solution, the planet was started at eight different values of the true anomaly. When one of those initial positions led to an unstable planetary orbit, then the specific ${(q_p,e_p)}$ pair was classified as unstable and we moved on to the next pair of values.  Additionally, we used the results of the numerical integrations performed for the nominal solutions in order to check whether or not the nominal solutions were close to the exact resonance, as defined in Murray \& Dermott (1999), Eq. (8.22).

In order to perform our numerical simulations, we used the symplectic integrator with time 
transformation developed by Mikkola (1997), specially designed to integrate hierarchical triple systems. 
The code uses standard Jacobi coordinates, i.e. it calculates the relative position and velocity 
vectors of the inner and outer orbit at every time step.  
Then, by using standard 
two body formulae, we computed the orbital elements of the two binaries.

All bodies were treated as point masses and only gravitational interactions at the Newtonian level were considered. Effects such as stellar finite size and general relativity were ignored at this point. The approximation of treating the bodies as
point masses, in particular for the inner binary, should be taken with some caution. Tidal deformation of the stars as well as general relativity effects might affect the dynamics of the system (e.g. see  Soderhelm 1984; Kiseleva, Eggleton \& Mikkola 1998, Eggleton, Kiseleva \& Hut 1998, Borkovits, Forg\'acs-Dajka \& Reg\'aly 2004).  These effects will be discussed in a the next section. 

\section{Results}
\subsection{Long term results}
First, we present the results from our long term numerical simulations.  As we stated previously, the nominal orbital
solutions of six Kepler circumbinary systems were integrated forward in time over 1 Gyr. During our simulations, none of the systems demonstrated any sign of instability for the initial conditions used.  
Figure 1 shows the variation of the planetary semi-major axis and eccentricity for the Kepler-64 system (Kostov et al. 2013 solution).  For this system,  the minimun and maximum semi-major axis values were 0.638 AU and 0.657 AU respectively, while for the eccentricity, the minimum and maximum values were 0.06 and 0.18.  Therefore no significant excursions in semi-major axis and eccentricity that might lead to instability are noticed in the two plots.  
\subsection{Critical semi-major axis}
Besides checking the long term stability of the six circumbinary systems, we also found the critical planetary semi-major axis for
each system with the method described in the previous section.  In addition, we computed the critical semi-major axis for the six Kepler systems under investigation by using the results of some earlier work: the empirical formula of Holman \& Wiegert (1999) and the semi-analytical stability criterion of Mardling \& Aarseth (2001). 
\begin{figure}
\includegraphics[width=90mm,height=90mm,angle=0]{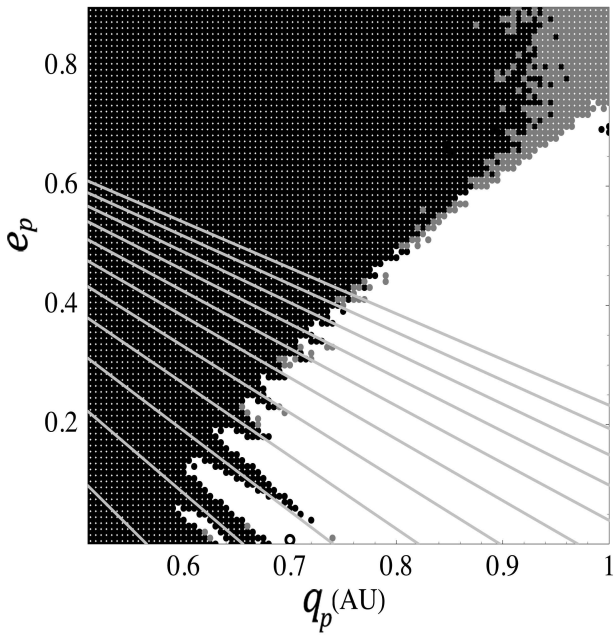}
\\
\caption {Eccentricity e$_p$ against pericentre distance q$_p$ for Kepler-16b. The integration time is 10$^{5}$yrs for the black dots and 10$^{6}$yrs for the dark grey dots. The empty circle (located rougly on the $q_p$ axis at the 0.7 AU) is the nominal position of the planet, and the light grey lines correspond to the locations of certain mean motion resonances between the stellar and planetary orbits.  From left to right, the resonances shown here are: 4:1, 5:1, 6:1, 7:1, 8:1, 9:1, 10:1, 11:1, 12:1, 13:1 and 14:1.}
\label{fig6}
\end{figure}
\begin{figure}
\includegraphics[width=90mm,height=90mm,angle=0]{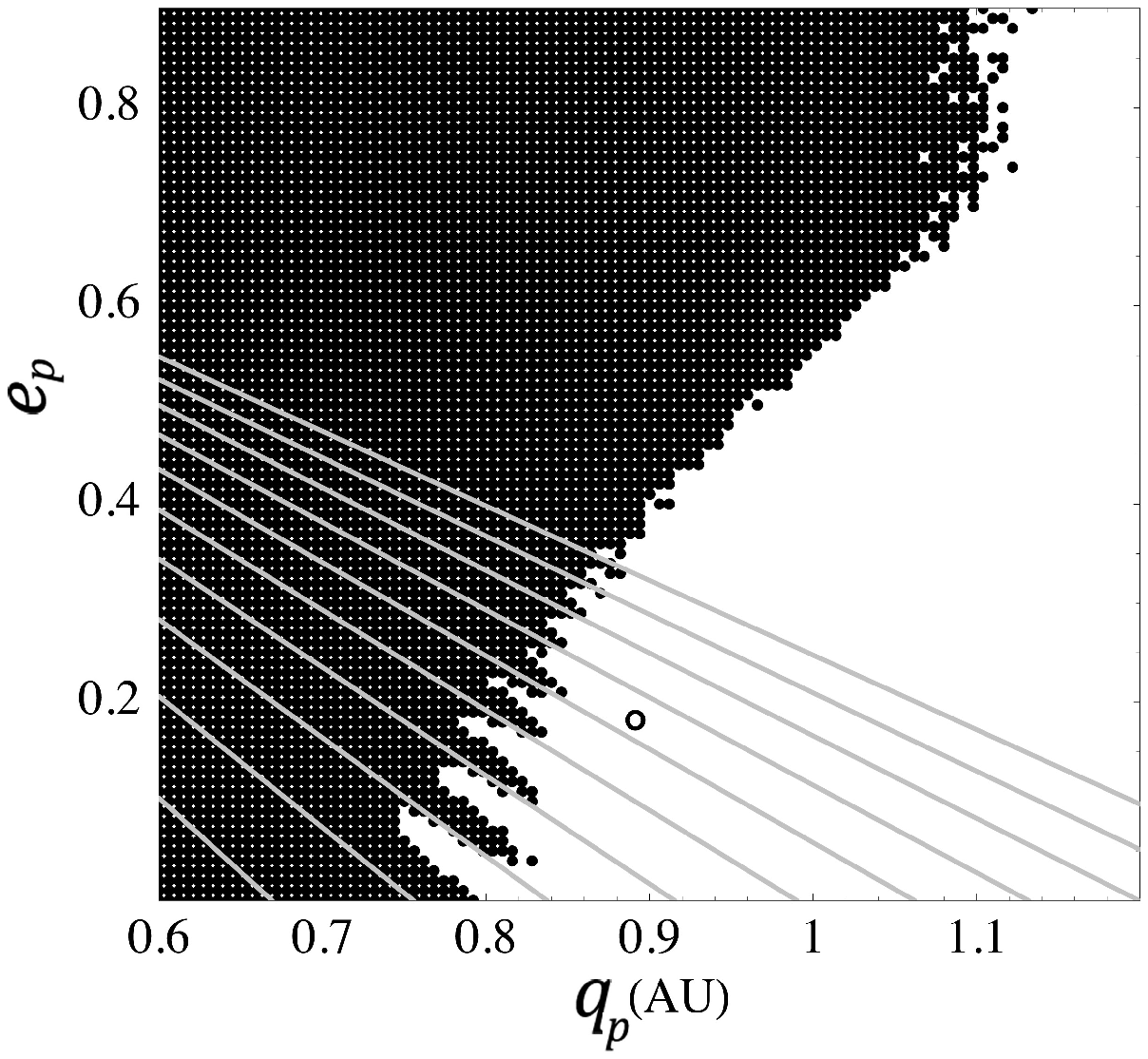}
\\
\caption {Eccentricity e$_p$ against pericentre distance q$_p$ for Kepler-34b. The integration time is 10$^{5}$yrs. The empty circle is the nominal position of the planet, and the light-grey lines correspond to the locations of certain mean motion resonances between the stellar and planetary orbits.  From left to right the resonances shown here are: 5:1, 6:1, 7:1, 8:1, 9:1, 10:1, 11:1, 12:1, 13:1 and 14:1.}
\label{fig7}
\end{figure}
\begin{figure}
\includegraphics[width=90mm,height=90mm,angle=0]{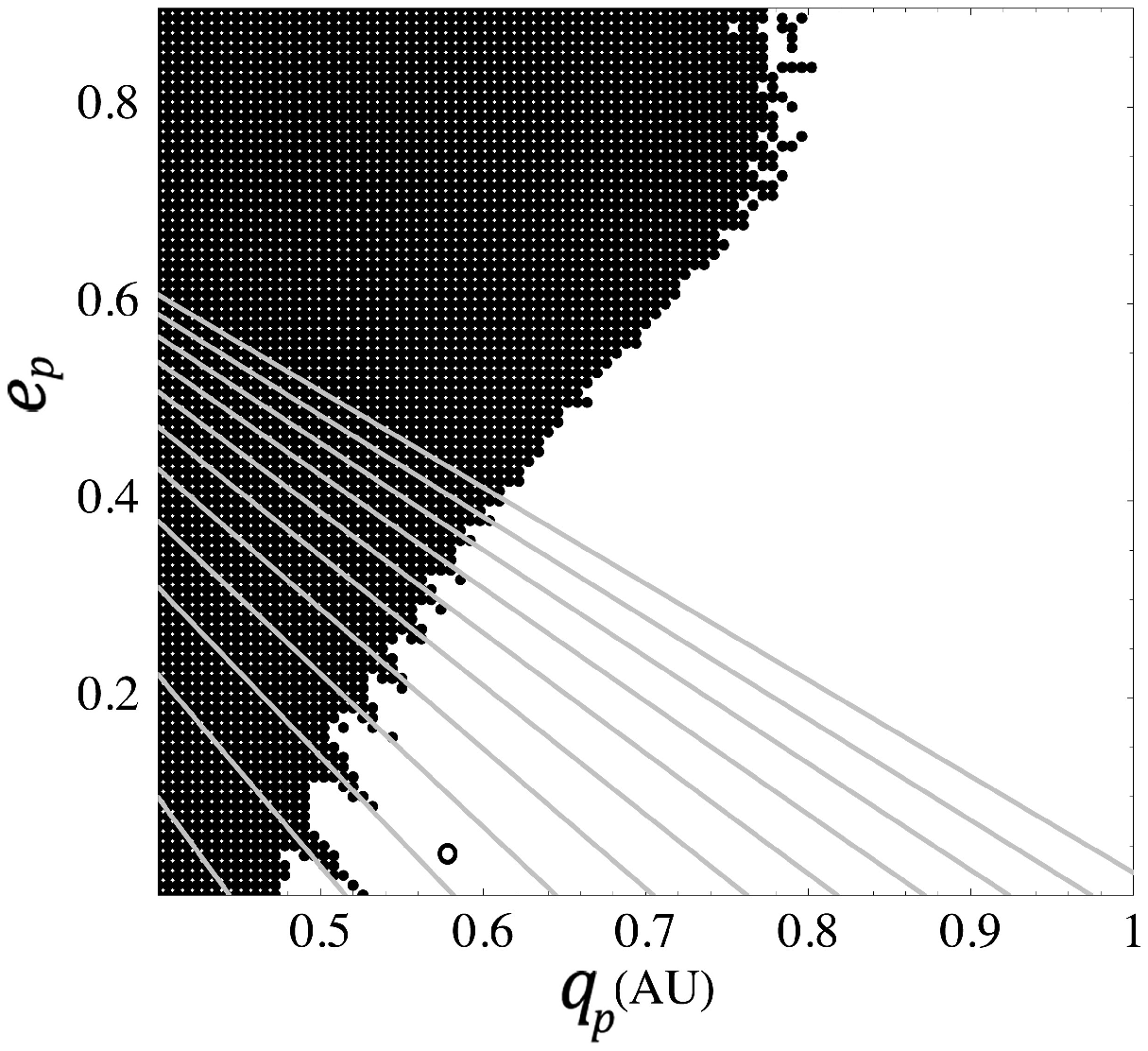}
\\
\caption {Eccentricity e$_p$ against pericentre distance q$_p$ for Kepler-35b. The integration time is 10$^{5}$yrs. The empty circle is the nominal position of the planet, and the light-grey lines correspond to the locations of certain mean motion resonances between the stellar and planetary orbits.  From left to right the resonances shown here are: 4:1, 5:1, 6:1, 7:1, 8:1, 9:1, 10:1, 11:1, 12:1, 13:1 and 14:1.}
\label{fig8}
\end{figure}
\begin{figure}
\includegraphics[width=90mm,height=90mm,angle=0]{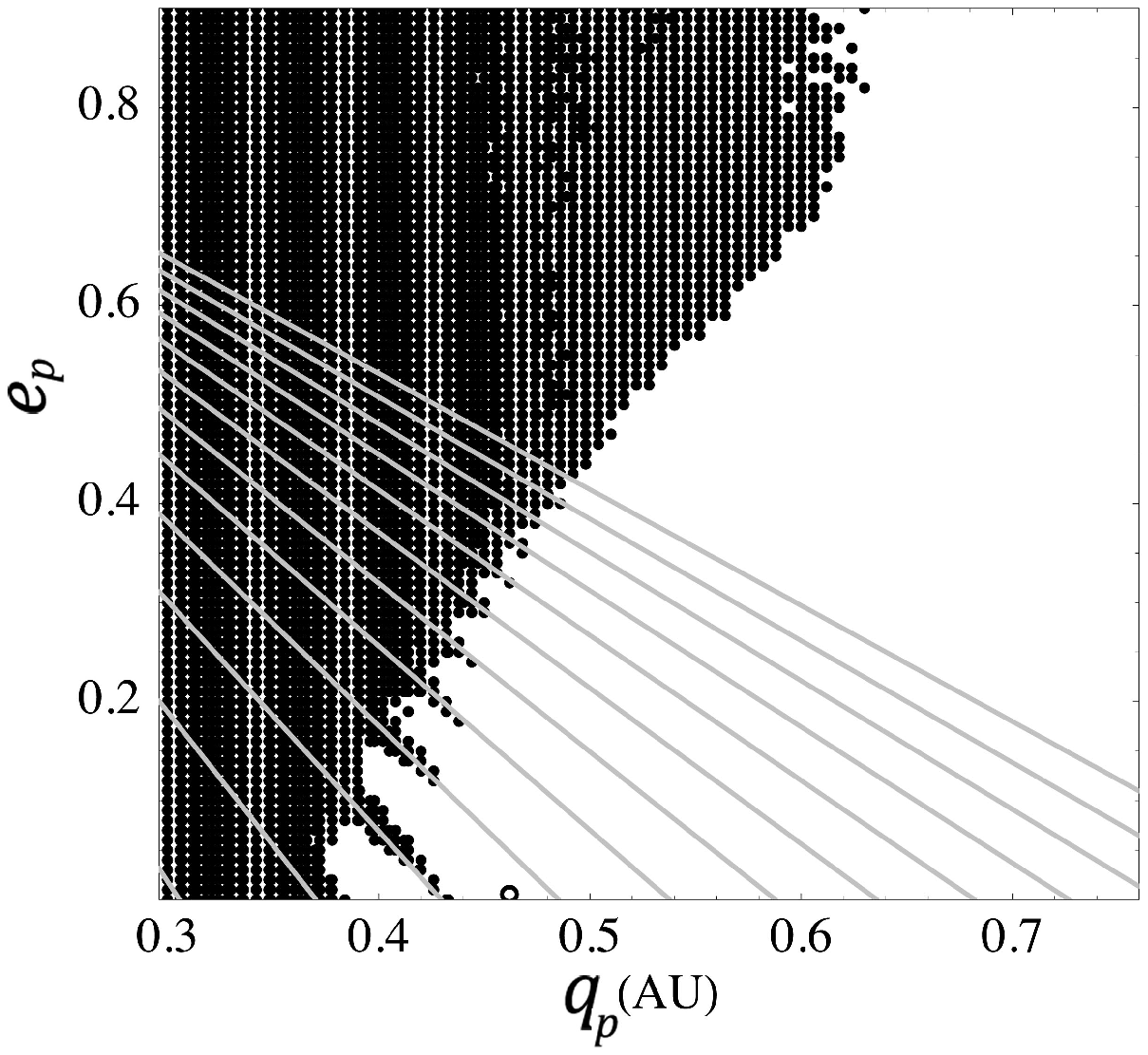}
\\
\caption {Eccentricity e$_p$ against pericentre distance q$_p$ for Kepler-38b. The integration time is 10$^{5}$yrs. The empty circle is the nominal position of the planet, while the light-grey lines correspond to the locations of certain mean motion resonances between the stellar and planetary orbits.  From left to right the resonances shown here are: 3:1, 4:1, 5:1, 6:1, 7:1, 8:1, 9:1, 10:1, 11:1, 12:1, 13:1 and 14:1.}
\label{fig9}
\end{figure}
\begin{figure}
\includegraphics[width=90mm,height=90mm,angle=0]{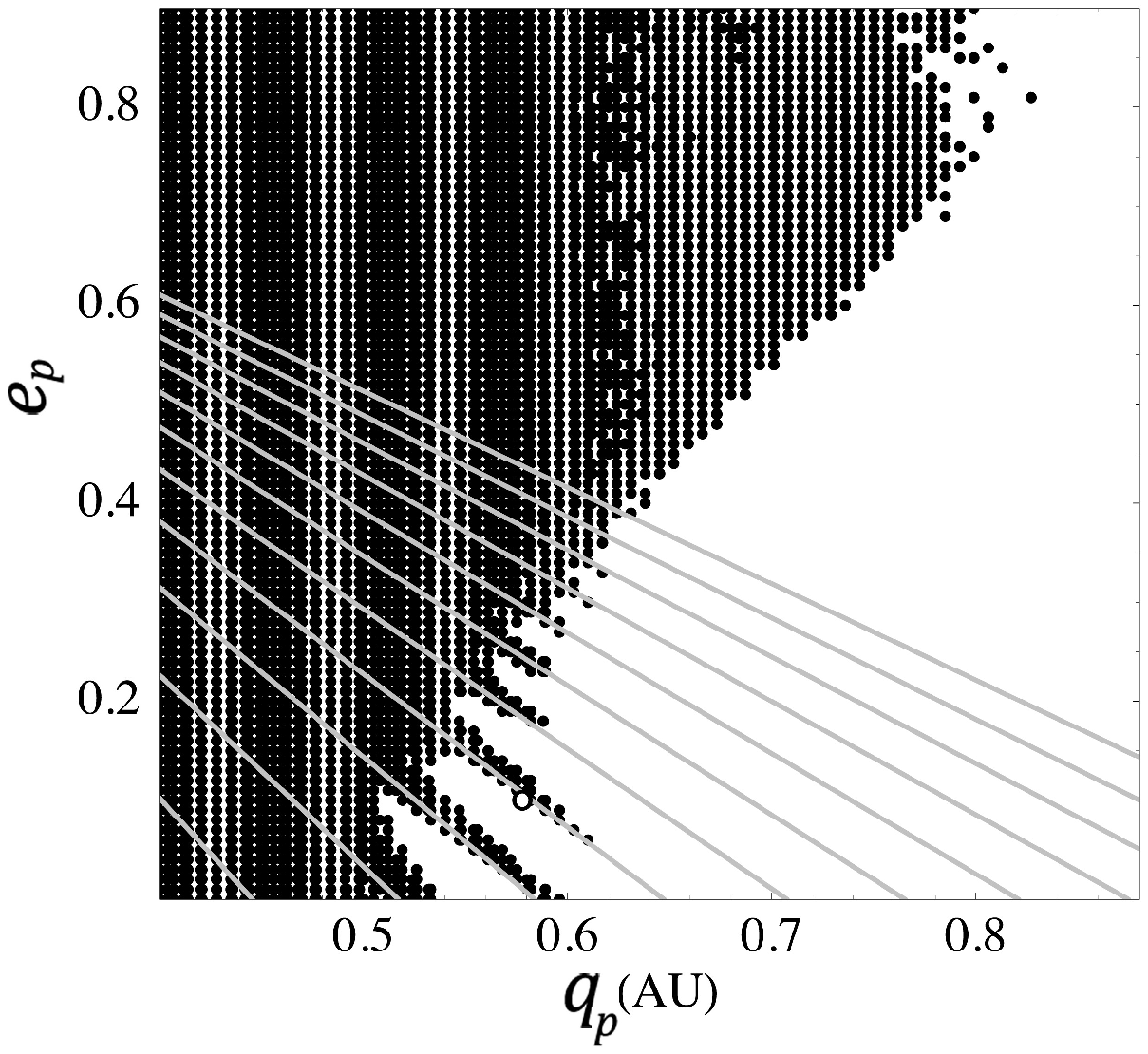}
\\
\caption {Eccentricity e$_p$ against pericentre distance q$_p$ for Kepler-64b based on Kostov et al. (2013). The integration time is 10$^{5}$yrs. The empty circle is the nominal position of the planet, while the light-grey lines correspond to the locations of certain mean motion resonances between the stellar and planetary orbits.  From left to right the resonances shown here are: 4:1, 5:1, 6:1, 7:1, 8:1, 9:1, 10:1, 11:1, 12:1, 13:1 and 14:1.}
\label{fig10}
\end{figure}
\begin{figure}
\includegraphics[width=90mm,height=90mm,angle=0]{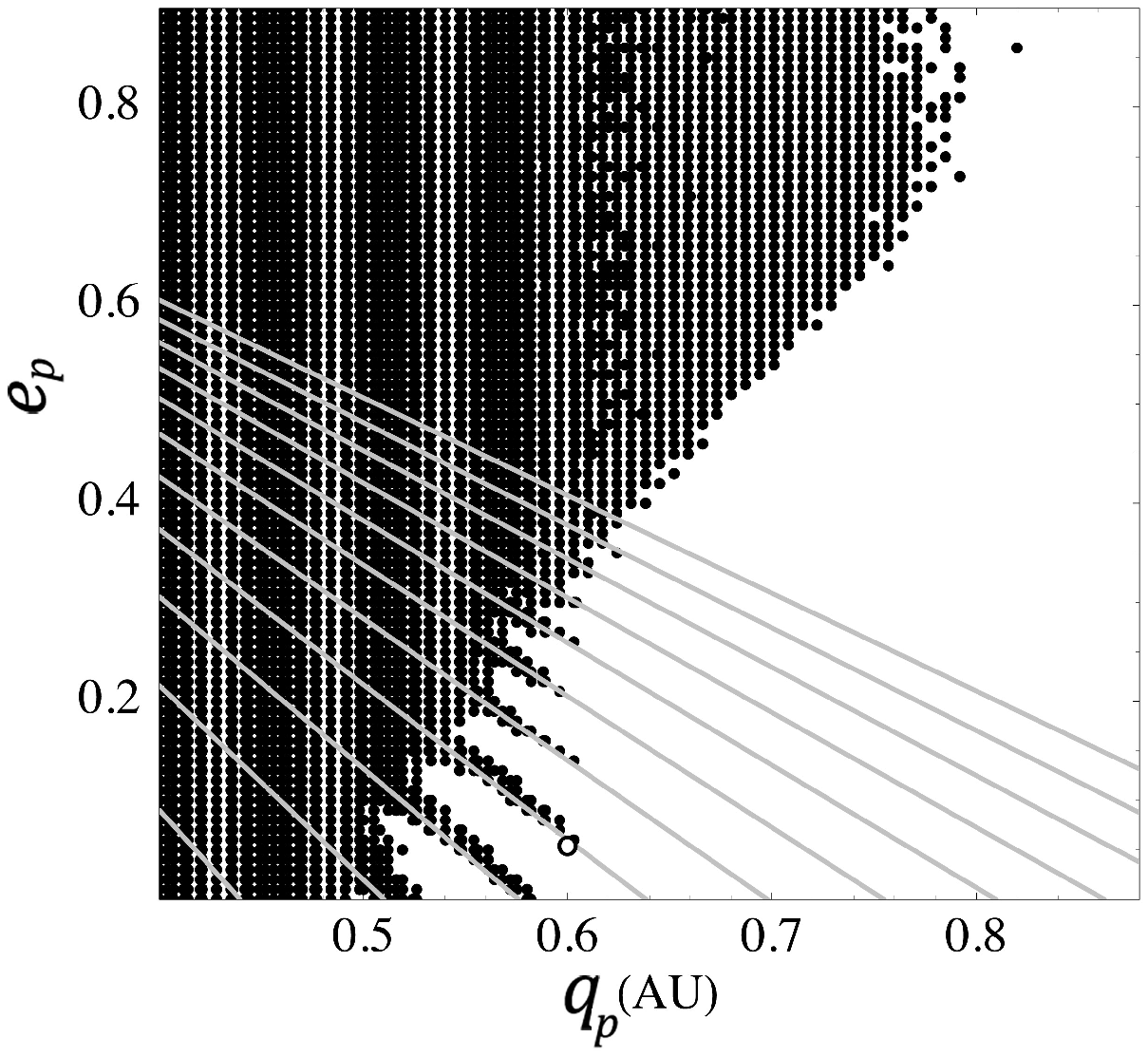}
\\
\caption {Eccentricity e$_p$ against pericentre distance q$_p$ for Kepler-64b based on Schwamb et al. (2013). The integration time is 10$^{5}$yrs. The empty circle is the nominal position of the planet, while the light lines correspond to the locations of certain mean motion resonances between the stellar and planetary orbits.  From left to right the resonances shown here are: 4:1, 5:1, 6:1, 7:1, 8:1, 9:1, 10:1, 11:1, 12:1, 13:1 and 14:1.}
\label{fig11}
\end{figure}
\begin{figure}
\includegraphics[width=95mm,height=87mm,angle=0]{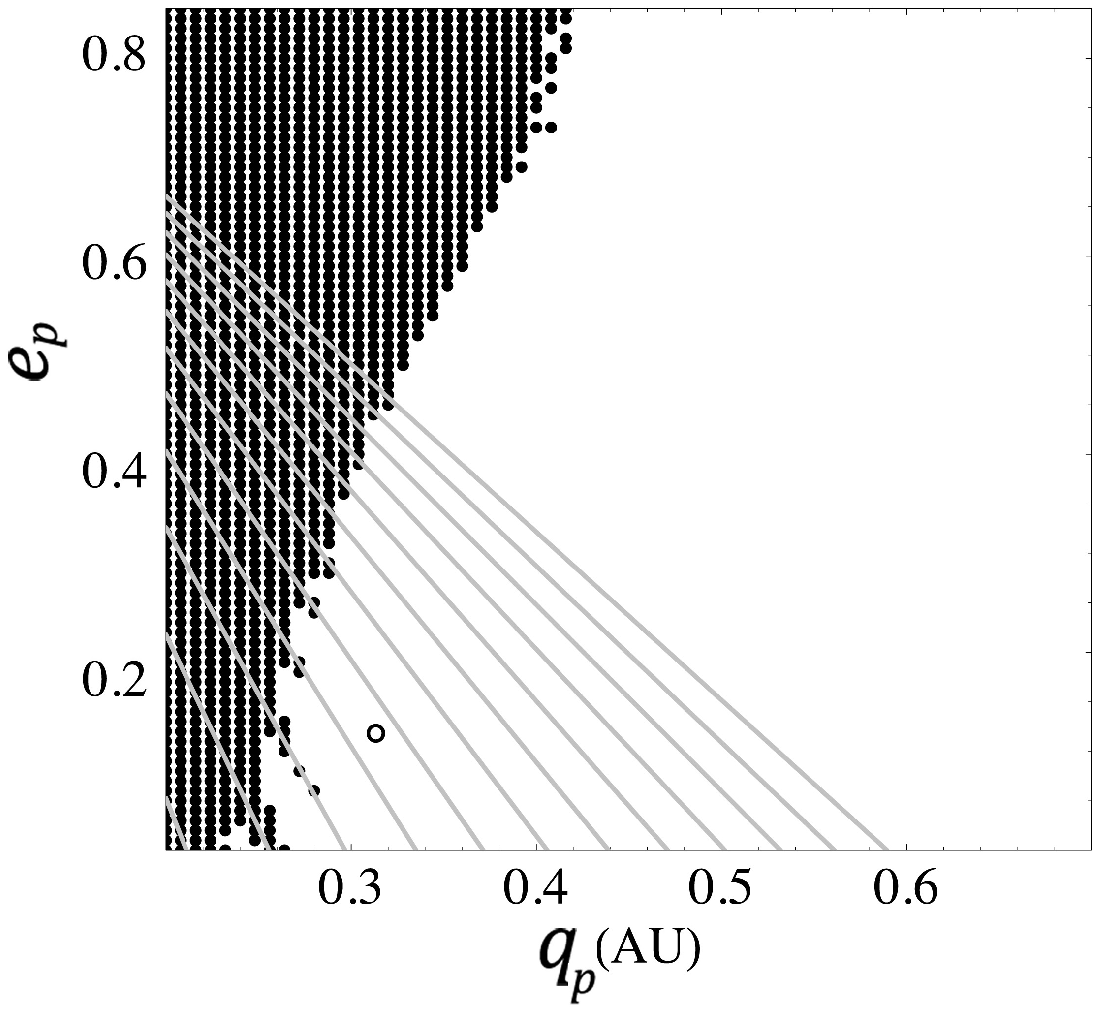}
\\
\caption {Eccentricity e$_p$ against pericentre distance q$_p$ for Kepler-413b. The integration time is 10$^{5}$yrs. The empty circle is the nominal position of the planet, while the light-grey lines correspond to the locations of certain mean motion resonances between the stellar and planetary orbits.  From left to right the resonances shown here are: 3:1, 4:1, 5:1, 6:1, 7:1, 8:1, 9:1, 10:1, 11:1, 12:1, 13:1 and 14:1.}
\label{fig12}
\end{figure}
Holman and Wiegert (1999) investigated  the stability of P-type and S-type orbits in stellar binary systems.
They performed numerical simulations of  particles  on initially circular and prograde circumstellar or circumbinary orbits, 
in the binary plane of motion and with different initial orbital longitudes.  The binary mass ratio was taken in the
range ${0.1 \leq \mu \leq 0.9}$, the binary eccentricity was in the range ${0.0 \leq e \leq 0.8}$ and the integrations
lasted for ${10^{4}}$ binary periods.  If a particle survived the whole integration time at all initial longitudes, then the system was classified as stable.  For circumbinary orbits, and by using a least squares fit to their data, they obtained an expression for the critical semimajor axis for the planet, $a_c$, given  by:
\begin{eqnarray}
a_{c} & = & [(1.60\pm 0.04)+(5.10\pm 0.05)e+(-2.22\pm 0.11)e^{2}+\nonumber\\
& & +(4.12\pm 0.09)\mu +(-4.27\pm 0.17)e\mu+(-5.09\pm \nonumber\\
& & \pm 0.11)\mu^{2}+(4.61\pm 0.36)e^{2}\mu^{2}]a_{b},
\label{hol2}
\end{eqnarray}
where ${a_{b}}$ is the binary semi-major axis, ${e}$ is the binary eccentricity and
$\mu={M_{2}/(M_{1}+M_{2})}$. 

Mardling and Aarseth (1999, 2001) approached the problem of stability of a hierarchical triple system by
noticing that stability against escape in the three body problem was analogous to stability
against chaotic energy exchange in the binary-tides problem. The way energy and angular
momentum are exchanged between the two orbits of a stable (unstable) hierarchical triple
system is similar to the way they are exchanged in a binary undergoing normal (chaotic) tide-orbit
interaction. Having that in mind, they derived a semi-analytical formula for the critical 
value of the outer pericentre distance ${R_{p}^{crit}}$:

\begin{equation}
\label{mard}
R_{p}^{crit}=C\left[(1+q_{out})\frac{1+e_{out}}{(1-e_{out})^
{\frac{1}{2}}}\right]^{\frac{2}{5}}a_{in},
\end{equation}
where  ${q_{out}=m_{p}/(M_{1}+M_{2})}$, ${e_{out}}$ is the outer binary eccentricity and ${a_{in}}$ is 
the inner semi-major axis. If
${R_{p}^{crit}\leq R_{p}^{out}}$, then the system is considered to
be stable.  The above formula is valid for prograde and coplanar
systems and it applies to escape of the outer body. The constant $C$ was
determined empirically and it was found to be 2.8. A
heuristic correction ${f=1-0.3i/180}$ (with ${i}$ being the mutual inclination in degrees) 
is applied for non-coplanar orbits, to account for the increased stability.  
Also, the criterion ignores  a weak dependence on the inner 
eccentricity  and inner mass ratio.
Finally, as stated in Mardling and Aarseth (2001), the formula holds for ${q_{out} \leq 5}$,
beyond which all unstable systems for the planets can suffer exchange; however the formula has not been tested for systems with planetary masses (Aarseth 2004), probably because the authors were mainly
interested in using the formula in star cluster simulations.
\begin{table*}\centering
 \caption{Critical planetary semi-major axis for Kepler-16, Kepler-34, Kepler-35, Kepler-38, Kepler-64 and Kepler 413.  'W',
'O', 'K13', 'S' and 'K14' stand for Welsh et al. (2012), Orosz et al. (2012a), Kostov et al. (2013), Schamb et al. (2013) and Kostov et al. (2014) respectively.} 
 \label{tab:2}
 \begin{tabular}{lrrrrrr}
    \hline
    \hline
    System & Nominal & Numerical & Holman \& Wiegert & Mardling \& Aarseth & Published\\
           &         &           &                   &                     &              \\
     & (AU) & (AU) & (AU) &  (AU) & (AU)  \\
    \hline
    \hline
    Kepler-16 & 0.7048     & 0.67   & 0.65 & 0.64  & 0.59 (W) \\
    Kepler-34 & 1.0896     & 1.00   & 0.84 & 0.87  & 0.88 (W)\\
    Kepler-35 & 0.60345    & 0.52   & 0.50 & 0.53  & 0.49 (W)\\
    Kepler-38 & 0.4644     & 0.43   & 0.39 & 0.43  &0.37 (O)\\
    Kepler-64 (K13) & 0.642  & 0.65  & 0.53 & 0.58  &$-$ (K13)\\
    Kepler-64 (S) & 0.634  & 0.58   & 0.52 & 0.53  & 0.57 (S) \\
    Kepler-413  & 0.3553  & 0.31 & 0.26 & 0.35  & $-$ (K14) \\
       \hline

  \end{tabular}

\end{table*}

The results regarding the critical semi-major axis can be found in Table 3, which, besides the values given by the two formulae just mentioned, also presents some relevant results by Welsh et al. (2012), Orosz et al. (2012a) and Schwamb et al. (2013).  

Generally, our simulations gave us a larger critical semi-major axis than all previous results in all but one case (for Kepler-413 the Mardling \& Aarseth criterion leads to a critical semimajor axis greater than our value). However, this does not necessarily 
mean that our results are inconsistent with those of the other authors.  The main reason for that is the definition of the
critical semi-major axis: we found the semi-major axis for which the system was stable and immediately after that the system became
unstable (although there may also be smaller values of the semi-major axis for which the system becomes stable again), while for example
Holman and Wiegert (1999) looked for the smallest semi-major axis for which the system was stable.  The definition of the critical axis in Welsh et al. (2012) is also similar to that. It is worth pointing out that for Kepler-16 Jaime et al. (2012), using the invariant loops criterion of Pichardo et al. (2005, 2008), found a value of 0.63 AU. Again, the value we found for this system (0.67 AU) is greater than that.

Another factor that may explain part of the discrepancy between our value for the critical semi-major axis and the rest of the results is the definition of stability on which those results were based. It is a factor that may have a significant effect on any such study (e.g. see Georgakarakos 2008).   For instance, the stability criterion in Welsh et al. (2012) would be expected to produce more conservative estimates.  However, the fact that their integration time was only the one tenth of our simulation time could possibly cancel out
that effect.  Similarly, Holman's and Wiegert's integration time is only a fraction of ours, but as we said, in that case, it is the
definition of the critical axis that explains the discrepancy between the results.

Finally, our results for Kepler-16, Kepler-34 and Kepler-35 seem to be in agreeement with the results obtained from Popova \& Shevchenko (2013).

\subsection{Stability maps}

In this subsection, we present the results from the third numerical experiment we conducted, which was the construction of stability maps for all planetary
systems under investigation.  In the way described in section 2, we found stable and unstable areas in the ${(a_p,e_p)}$ plane for
each system and the results can be seen in Figures 2-8.  The areas denoted by black dots show the unstable planetary orbits, while the white area represents the stable ones.  The location of the nominal solution for each system is denoted by an empty black circle
and also, indicated by straight grey lines, we mark the location of various mean motion resonances between the binary system and the planet. The resonances are of the type k:1, with k ranging from 4 to 14 for Kepler-16, from 5 to 14 for Kepler-34, from 4 to 14 for Kepler-35, from 3 to 14 for Kepler-38, from 4 to 14 for Kepler-64 (both solutions) and from 3 to 14 for Kepler-413.
As we stated earlier, in this experiment, each point on the grid was integrated for ${10^5}$ years.  Since we deal with systems that consist of two stars and a planet, a stability simulation may require a very long integration timescale,
as sometimes instability signs may not become evident fast, but they become noticeable after a significant number
of secular periods (e.g. see Georgakarakos 2013).  However, for the six Kepler systems, the planetary secular periods of motion is of the order of a few decades and therefore our ${10^5}$ year integration time should be sufficient for investigating the stability of the systems.

Fig. \ref{fig6} shows the stability map of the Kepler-16 system.  The planet seems to reside in a stable area, close to a region of
unstable orbits associated with the 5:1 and 6:1 mean motion resonances.  Our map also is consistent with the corresponding one of Popova \& Shevchenko (2013).  There are some areas in the map denoted by light grey dots. These correspond to unstable planetary orbits which were integrated for 1 Myr, ten times longer than the rest of the simulations.  This was done in order to get an idea of how a longer integration time could possibly affect our stability results and maps. Apparently, only a limited area was added to the map, mainly for higher eccentricity values.

Figure \ref{fig7} shows the stability map for Kepler-34b.  The planet seems to be far enough from the unstable area,
lying between the 10:1 and 11:1 mean motion resonances.  The unstable area in our map is larger than the corresponding plot of Popova and Shevchenko (2013), probably due to the longer integration time we used.

Regarding Kepler-35, as seen from Fig. \ref{fig8}, the planet is located in the stable area on the ${(a_p,e_p)}$ plane, between
the 6:1 and 7:1 resonances.  As before, our map has a more extensive unstable area than the corresponding map of Popova and Shevchenko (2013), especially for intermediate and high eccentricities.

The Kepler-38 stability map is shown in Fig. \ref{fig9}.  The planet is clearly sitting in a stable area between two prongs of unstable planetary orbits which correspond to the 5:1 and 6:1 mean motion resonances.

For the Kepler-64 system we have two figures, Figure \ref{fig10}, which is based on the solution of Kostov et al. (2013) and Figure \ref{fig11},which is based on the solution of Schwamb et al. (2013).  For both models, the nominal solution for the planetary orbit is 
very close to the unstable points associated with the 7:1 mean motion resonance.  However, the stability maps have been based
on an upper limit for the planetary mass.  It is possible that a smaller planetary mass would result in a different stability map (e.g. Georgakarakos 2013).   We would like to point out here that, as it is most likely, the planet did not form at its present location, which means that it possibly had to go through areas of instability during its migration phase.  However, the crossing of the planet through the unstable areas during its inward migration phase would still be possible due to the facts that the migration timescale could be shorter than the instability timescale and that eccentricity damping forces can lower the planetary eccentricity during that phase.  Of course, the previous remark is valid not only for Kepler-64, but also for the rest of the systems we investigate here.

Finally, the last planetary system that we model here is Kepler-413.  Fig. \ref{fig12} shows the stability map for the system.  The orbit of the planet is lying in a stable area between the 6:1 and 7:1 mean motion resonances.

\subsection{Checking if the nominal solutions are in resonance}
\begin{figure}
\includegraphics[width=90mm,height=60mm,angle=0]{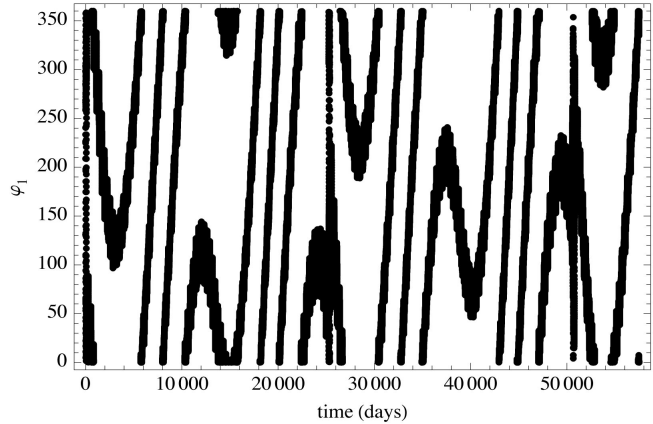}
\caption {Time evolution of the resonant angle $\phi_{1}= \lambda_{b} -10 \lambda_{p} + 9 \varpi_{b}$ for Kepler-34b.  Both circulation and libration features can be noticed in the above plot,
which is a typical behaviour for motion near the separatrix between libration and circulation, indicating that the system is close to the exact resonance.}
\label{fig57}
\end{figure}

\begin{figure}
\includegraphics[width=90mm,height=60mm,angle=0]{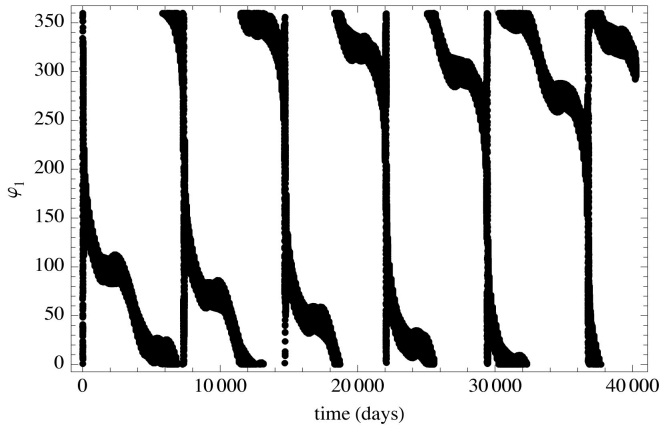}
\caption {Time evolution of the resonant angle $\phi_{1}= \lambda_{b} -6 \lambda_{p} + 5 \varpi_{b}$  for Kepler-35b.  Again, as in the case of Kepler-34b, though not in the same degree, both circulation and libration features can be noticed in the above plot.}
\label{fig58}
\end{figure}

As we saw earlier, all planets investigated here are close to a mean motion commensurability with the stellar binary and therefore we decided to check whether the systems were close to the exact resonance, using the definition as it appears in Murray and Dermott (1999) subsection 8.9.1.  For that reason, we explored all possible 
resonant angles.  For example, Kepler-16b sits between the 5:1 and 6:1 MMR and hence, in this case, the potentially resonant angles are given by $\phi= \lambda_{b} -5 \lambda_{p} +j_3 \varpi_{b} + j_4 \varpi_{p}$ for the 5:1 MMR or $\phi= \lambda_{b}-6 \lambda_{p} +j_{3} \varpi_{b} + j_{4} \varpi_{p}$ for the 6:1 MMR, where $\lambda_{b}$ is the mean longitude of the stellar binary, $\lambda_{p}$ is the mean longitude of the planet, $\varpi_{b}$ is the longitude of the pericentre of the stellar binary and  $\varpi_{p}$ is the longitude of the pericentre of the planet. The integer numbers $ j_{3}$ and $j_{4}$ must satisfy the relationship $1-5+j_{3}+j_{4}=0$ for the 5:1 MMR and $1-6+j_{3}+j_{4}=0$ for the 6:1 MMR (Murray \& Dermott 1999).
We found that all the possible resonant angles of the systems were circulating at a high frequency and therefore not close to the exact resonance.  The two most interesting exceptions were Kepler-34 and Kepler-35.  Regarding Kepler-34, although the planet is closer to the 10:1 MMR than to the 11:1, we checked both possibilities. We found that the nominal solution is close to the exact resonance associated with the 10:1 MMR. Fig. \ref{fig57} shows the time evolution of the resonant angle $\phi_{1}= \lambda_{b} -10 \lambda_{p} + 9 \varpi_{b}$.  Both circulation and libration features can be noticed in that figure.  Similarly, for Kepler-35, which is between the 6:1 MMR and the 7:1 MMR, we found that the important angle was $\phi_{1}= \lambda_{b} -6 \lambda_{p} + 5 \varpi_{b}$.  Its time evolution can be seen in Fig. \ref{fig58}. Again, the plot exhibits both circulation and libration features.
\begin{figure}
\includegraphics[width=75mm,height=70mm,angle=0]{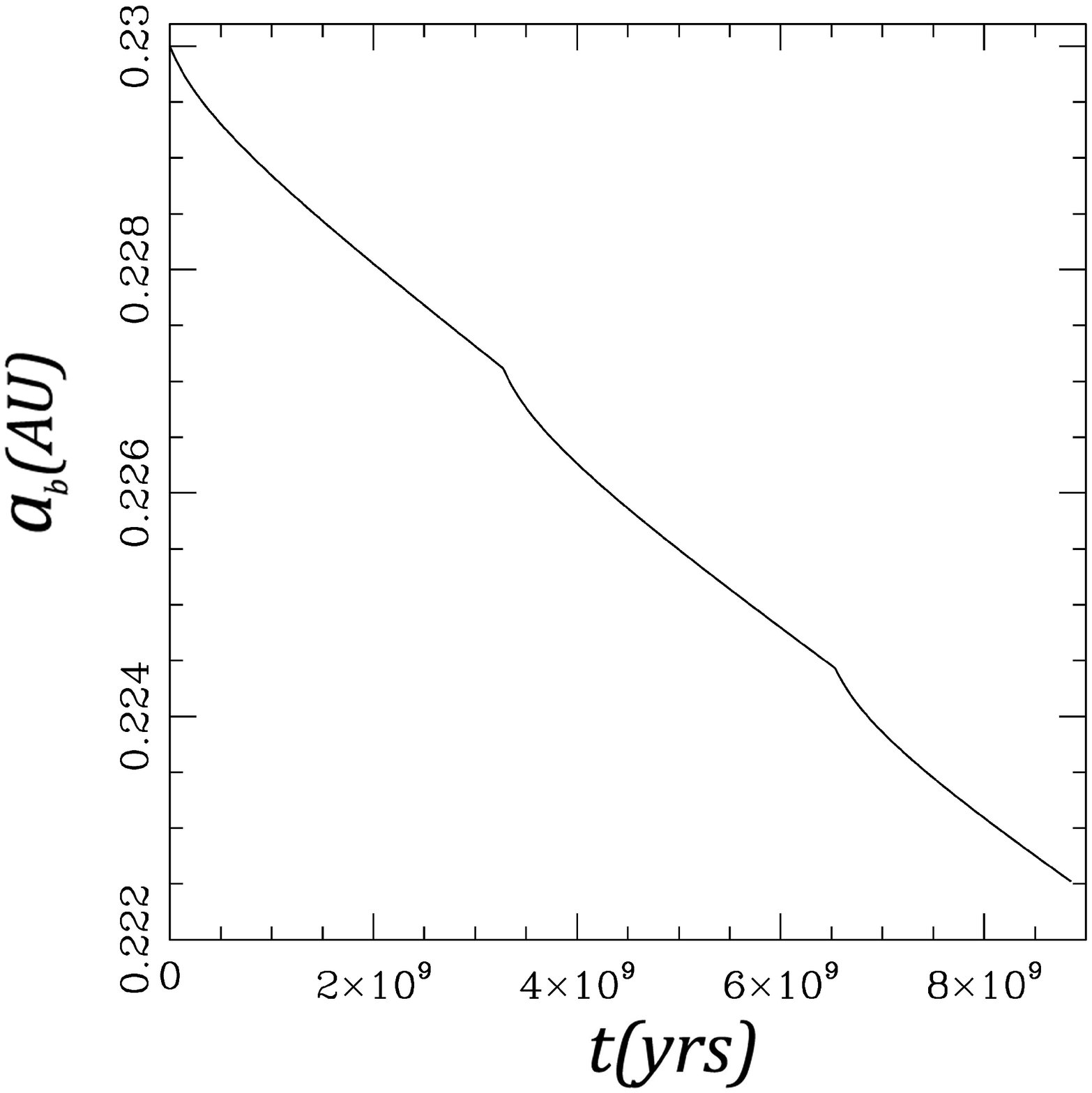}
\includegraphics[width=75mm,height=70mm,angle=0]{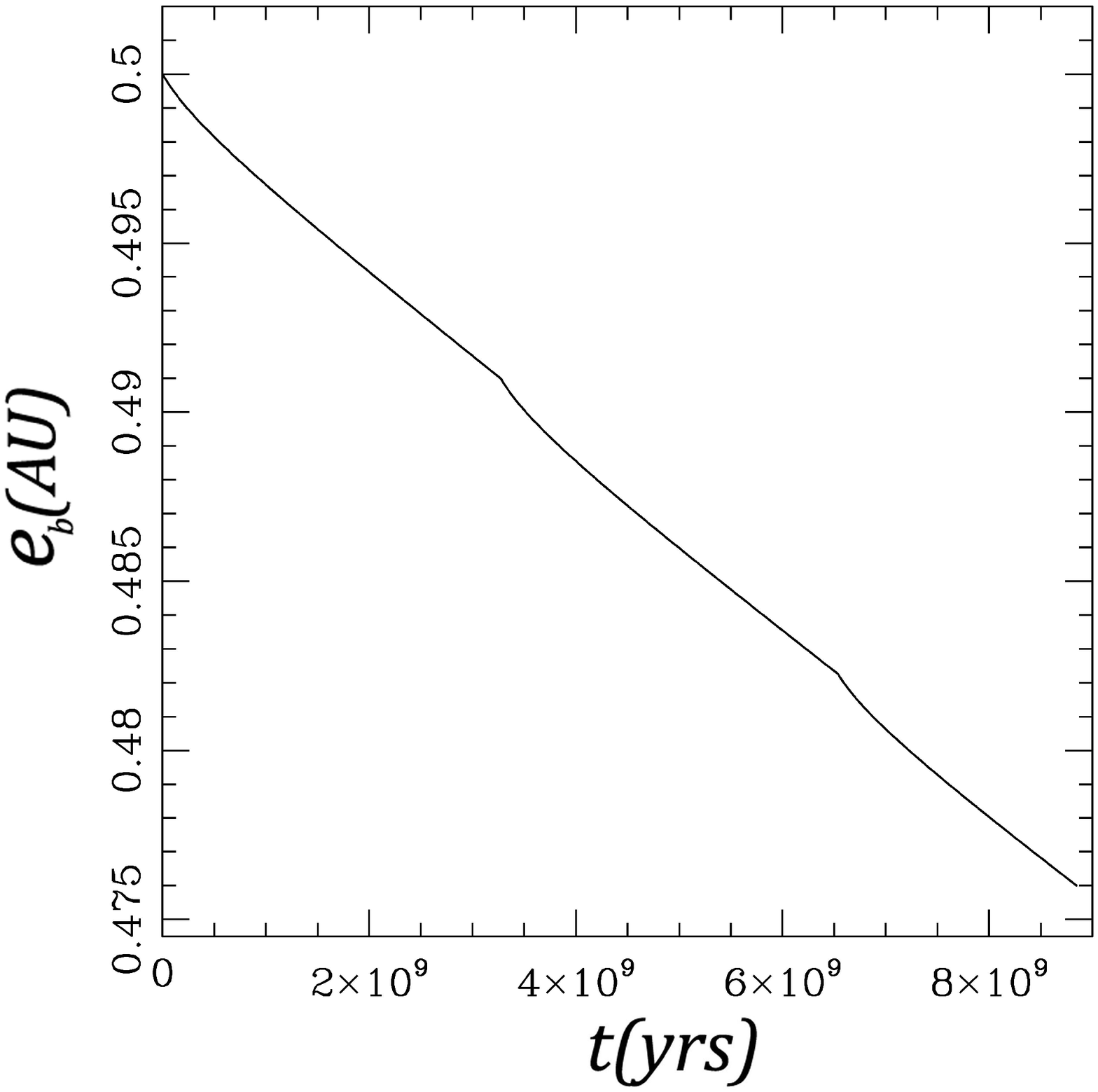}\\
\caption {Semi-major axis and eccentricity of the inner binary orbit for the Kepler-34 system under tidal, star deformation due to rotation and general relativity effects. 
The integration time is 9$\times$10$^{9}$yrs. }
\label{fig13}
\end{figure}
\section{Discussion and Summary}
\subsection{Effect of tidal evolution and general relativity}

As it was stated earlier, the previous simulations were done under the assumption
that the bodies were point masses. However, when the binary separation is
of the order of a few stellar radii, tidal friction may become important, particularly if a system undergoes evolutionary phases of 
high eccentricity. Tidal dissipation can affect the orbital evolution of the stellar binary, which in turn, may have a significant
effect on the planetary orbit. In addition, the orbit may also be affected by the non-spherical shape that the stars acquire due to the tidal effects and due to their rotation around their axes.  Finally, general relativistic effects may also play a role in the stellar binary orbital evolution, as the extra frequency due to the precession of the longitude of the pericentre may have an effect on the orbital evolution.

From the systems we considered, Kepler-34 experiences the closest pericentre distance between the stellar components in terms of their stellar diameter and for that reason it is chosen to see whether the above mentioned effects may become important in the evolution of the system (Kepler-34 is the largest mass-combined system with 2.0687 M$_\odot$ and has the largest eccentricity, both important factors for GR).  More specifically, ${e_{b}=0.52087}$ and ${a_{b}=0.22882 AU}$, which yield a pericentre distance 
of 0.10963 AU, which is approximately ten times the diameter of the largest star of the system. For comparison, the Kepler-413 has ${e_{b}=0.0365}$ and ${a_{b}=0.10148AU}$ that give us a pericentre distance of 0.0978 AU, which is approximately 13.5 times the diameter of the largest star on the system. Additionally, the combined mass of the stars in Kepler-413 is 1.3623 M$_\odot$ 

We integrated numerically the equations of motion (Eggleton et al. 1998, Prodan \& Murray 2012) by
using a Bulirsch-Stoer integrator (Press et al. 1996). 
The initial conditions used were: semi-major ${a_{b}=0.23AU}$, eccentricity ${e_{b}=0.5}$, longitude of pericentre ${\varpi_b=0^\circ}$, $m_{1}=1.05$ $M_{\odot}$, $m_{2}=1.02 M_{\odot}$, radius of the first star ${R_{1}=1.16R_{\odot}}$, radius of the second star ${R_{2}=1.09R_{\odot}}$, spin period for both stars 16 days, tidal Love number
${k_{2}=0.028}$ and tidal dissipation factor ${Q=10^{6}}$ (the last two are adopted values for sunlike stars). The presence of the planet was neglected and the integration time was set to 9 Gyrs.

As expected, the results showed that the binary orbit tended to shrink and to get circularised, i.e. there was a clear decrease in the values of the semi-major axis and eccentricity. The eccenticity reduces by about ${5\%}$ after 8 Gyrs, while the semi major axis exhibits a drop of around ${3\%}$ in the same period of time.  Finally, the argument of pericentre circulates with a period of around ${2\times10^{5}}$ yrs.  A graphical representation of some of the results can be found in Fig. \ref{fig13}.  Since the above timescales are much longer than the secular ones,  the evolution of the planetary orbit should not be significantly affected.

Hence, based on our results of Kepler-34, we conclude that tidal friction, general relativistic effects and the deformation of the stars due to rotation are not very important in modelling the systems we are studying here.

\subsection{Summary}
We have investigated the stability of the six Kepler single-planet circumbinary systems, i.e. Kepler-16, Kepler-34, Kepler-35, Kepler-38, Kepler-64 and Kepler-413 by integrating numerically the full equations of motion of each system.The orbits were assumed to be coplanar and our investigation was split into three experiments.

In the first experiment, we  integrated the nominal solution of each system, as given in the corresponding papers, for 1Gyr, a timescale  that was much longer than all previous 
work by other authors ($\leq$ 10Myrs), except for the Kepler-64 system (Schwamb et al. 2013), for which the authors used a similar timescale to ours. Our simulations showed that all systems were stable for the time that were integrated.  

In the second numerical experiment we searched for the critical planetary semi-major axis (as defined earlier). The integration was done for ${10^5}$ years which was longer than any previous timescale for a similar experiment.  Besides comparison of our results with the corresponding results of Welsh et al. (2012) and Orosz et al. (2012), we also compared our values for the critical semi-major axis with results based on the stability criteria of Holman \& Wiegert (1999) and Mardling \& Aarseth (2001). Generally, our simulations gave larger values compared to the rest of the results by other authors.

In the third experiment, we constructed stability maps on the ${(a_p,e_p)}$ plane.  The simulations were done for ${10^5}$ years,
which was at least 10 times longer that some similar past results.  Moreover, this analysis was done for the first time for Kepler-38.  The simulations showed that all planets reside in stable
areas between instability prongs along the location of mean motion resonances.  However, the position of Kepler-64b is very close to unstable orbits assosiated with the 7:1 mean motion resonance.

Additionally, we decided to use the data from the numerical integrations we performed for the nominal solutions in order to check whether the systems were close to the exact resonance and we found that Kepler-34 and Kepler-35  were the most interesting cases in that respect.  This result may be related to the dynamical evolution 
of those two systems during their formation stage.  In the context of planet formation around single stars, numerical simulations 
of the process of gas driven migration suggest that, as the gas in a 
protoplanetary disk is dispersed, giant planets can be found locked 
in a multiresonant configuration (Masset \& Snellgrove 2001, 
Lee \& Peale 2002, Morbidelli \& Crida 2007, Morbidelli et al. 2007). 
Whether this resonance-locking occurs also in the gaseous protoplanetary
disks around binary stars, has been studied by Nelson 2006 and Pierens \& Nelson (2008, 2013).
Our results suggest that, for some reason, Kepler-34 and possibly Kepler-35 demonstrate some features of that resonance trapping. 
The fact that we do not see similar features in the orbital motion of the rest of the systems we investigated, may be related to 
the further evolution of the systems due to the interaction of the planets with a remnant planetesimal 
disk, which can drive the planets out of the resonant condition, as is believed 
to have happened in the solar system (Malhotra 1995, for a review see Levison 
et al. 2007). This again may suggest a different evolutionary 
path for Kepler-34 (and possibly Kepler-35) compared to the rest of the systems 
in our sample, an interesting issue to be studied in future contributions.

The current reasearch therefore builds up on previous
evidence that these systems are stable.  Whether this scenario
may change due to new discoveries such as more planetary
companions or revised orbital solutions, only more observations
will tell.  Our results may be important for issues such as habitability, as 
the determination of habitable zones or looking for Earth-like planets in 
the Kepler systems under investigation can not ignore the presence of the 
already known gas giant planets. In addition, our
results, and especially our stability diagrams, could also 
prove very helpful when investigating planet formation scenarios
and especially late stage formation, when Lunar-sized embryos develop 
into fully formed planets.  As the stability limit for a planetary or an even
smaller mass body in a circumbinary orbit appears to have a weak dependence on
the mass of the planetary body (e.g. Georgakarakos 2013), our stability diagrams
can provide useful information regarding where a planetary body may form under the
gravitational perturbations of the six Kepler stellar binaries of this study.

\section{Acknowledgements}
We thank the referee for the useful comments and corrections. CC would lile to thank to FIME, UANL and Mexican SEP program PROMEP for their financial support on the development of this research. MRR acknowledges support from UNAM-DGAPA-PAPIIT project 115413 . HA thanks CONACyT Research Project 179662 and UNAM-DGAPA project 108914.

\label{lastpage}

\end{document}